\documentclass[10pt,twocolumn,letterpaper]{IEEEtran}
\usepackage{amsmath,epsfig}
\usepackage{algorithm}
\usepackage{bm}
\usepackage{tikz}
\usepackage{graphicx}
\usetikzlibrary{positioning}
\usetikzlibrary{shapes.multipart}
\usepackage{pgfplots}
\usepackage[usenames,dvipsnames]{pstricks}
\usepackage{epsfig}
\usepackage{pst-grad} 
\usepackage{pst-plot} 
\usepackage{subfigure}
\usepackage[capitalize]{cleveref}

\usepackage{subfigure}
\usepackage[T1]{fontenc}
\usepackage{amsmath}
\usepackage{graphics}
\include{psfig}
\usepackage{epsfig}
\usepackage{array}
\usepackage{psfrag}
\usepackage{amssymb}
\usepackage{color}
\usepackage{pstricks,pst-node,pst-text,pst-3d,pst-plot}
\usepackage{cite}

\usepackage{ulem}
\definecolor{dgreen}{rgb}{0, 0.8, 0.1}

\begin{document}

\title{Single Channel MMWave FMCW Radar for 2D Target Localization}
\author{Shahrokh Hamidi$^*$\thanks{Shahrokh Hamidi is with the Faculty of Electrical and Computer
Engineering, University of Waterloo, 200 University Ave W, Waterloo, ON., Canada, N2L 3G1.
e-mail: \texttt{Shahrokh.Hamidi@uwaterloo.ca}.} \and and Safieddin Safavi-Naeini$^*$\thanks{Safieddin Safavi-Naeini is with the Faculty of Electrical and Computer Engineering, University of Waterloo, 200 University Ave W, Waterloo, ON., Canada, N2L 3G1.
e-mail: \texttt{
safavi@uwaterloo.ca}.}}

\maketitle
\begin{tikzpicture}[remember picture, overlay]
      \node[font=\small] at ([yshift=-1cm]current page.north)  {This paper has been accepted for publication in the IEEE $\rm 19^{th}$ International
Symposium on Antenna Technology and Applied Electromagnetics, 2021. \copyright IEEE};
\end{tikzpicture}

\begin{abstract}
In this paper, we present a 2D target localization method using two low-cost and compact Mellimeter Wave Frequency Modulated Continuous Wave (MMW-FMCW) radars.
To create a 2D map we exploit the bilateration method followed by a multi-target tracking block to remove the ghost targets.

Finally, we present experimental results based on the data gathered from two FMCW radars operating at $\rm 79 \;GHz$.
\end{abstract}


\section{Introduction}
Millimeter Wave Frequency Modulated Continuous Wave (MMW-FMCW) radars have become very popular recently. They are low cost and compact sensors with low peak to power ratio. One single channel FMCW radar is capable of measuring the radial distance as well as the velocity of the targets located in the field of view of the radar. In order to measure the Angle Of Arrival (AOA), MMW-FMCW radars have been combined together to create sparse arrays in Multiple Input Multiple Output (MIMO) configurations which will make the AOA estimation possible \cite{Shahrokh, 24GHzTDM, 77GHzTDM, TDMcalibration, FDMMIMO, TDMmotioncompensation}.

However, MIMO radars have their own shortcomings. To obtain a satisfactory angular resolution limit, the number of elements should be high which in turn makes the system more complex. Issues such as heat sink, power consumption, antenna coupling, synchronization and calibration have to be considered when we increase the number of elements.

In this paper, we use two single channel MMW-FMCW radars that operate independently to localize the energy of the targets in 2D plane. We exploit the bilateration method to find the x-y position of the targets. In order to create two orthogonal signals in space the radars will operate in Time Division Multiplexing (TDM) configuration. Each MMW-FMCW radar is a small and compact single channel sensor with on-chip antennas.

The organization of the paper is as follows. In Section \ref{Model Description}, we present the system model and the entire signal processing and data processing blocks that are necessary for target localization based on the bilateration method. In Section \ref{Experimental Results}, we use experimental data gathered from two single channel MMW-FMCW radars operating at $\rm 79\;GHz$ and show the result of the 2D target localization based on the algorithm which we develop in the paper and discuss the results.
\section{Model Description}\label{Model Description}
To localize the targets based on bilateration method, we use two single channel FMCW radars. Each single channel radar provides the range profile of the targets located in its field of view. In fact, the output of the radar at the Analog to Digital Convertor (ADC) level is the beat signal. After taking Fourier transform from the beat signal we will obtain the range profile of the targets.

We will then apply the Constant False Alarm Rate (CFAR) adaptive threshold \cite{Mahafza} to the signal in order to detect the location of the targets. The CFAR block will extract the energy of the existing targets from the additive noise. Since the CFAR block outputs multiple points per each target, therefore, we exploit pruning technique to select the point with highest intensity per each target.
\begin{figure}[htb]
\centering
\begin{tikzpicture}
  \node (img1)  {\includegraphics[scale=0.7]{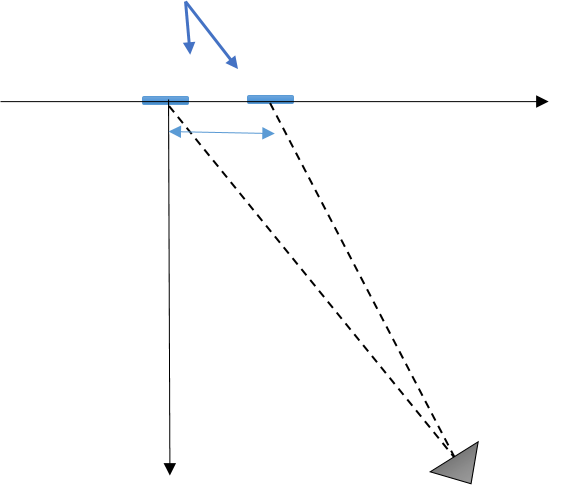}};
  \node[above=of img1, node distance=0cm, xshift=-0.7cm, yshift=-1.2cm,font=\color{black}] {The MMW-FMCW radars};
  \node[above=of img1, node distance=0cm, anchor=center,xshift=-1.3cm, yshift=-2.1cm,font=\color{black}] {\small Radar1};
  \node[above=of img1, node distance=0cm, anchor=center,xshift=3cm, yshift=-2.5cm,font=\color{black}] {\small x};
  \node[above=of img1, node distance=0cm, anchor=center,xshift=-1.1cm, yshift=-6.7cm,font=\color{black}] {\small y};
  \node[above=of img1, node distance=0cm, anchor=center,xshift=0cm, yshift=-2.1cm,font=\color{black}] {\small Radar2};
   \node[above=of img1, node distance=0cm, anchor=center,xshift=-0.7cm, yshift=-2.5cm,font=\color{black}] {\small d};
     \node[above=of img1, node distance=0cm, anchor=center,xshift=0.1cm, yshift=-4.5cm,font=\color{black}] {\tiny $R_1$};
       \node[above=of img1, node distance=0cm, anchor=center,xshift=1.2cm, yshift=-4.5cm,font=\color{black}] {\tiny $R_2$};
  \node[above=of img1, node distance=0cm, anchor=center,xshift=2.8cm, yshift=-6.3cm,font=\color{black}] {\small Target};
  \node[above=of img1, node distance=0cm, anchor=center,xshift=2.8cm, yshift=-6.6cm,font=\color{black}] {\tiny $(x_t,y_t)$};
\end{tikzpicture}
\caption{The system model.}
\label{fig:Model}
\end{figure}
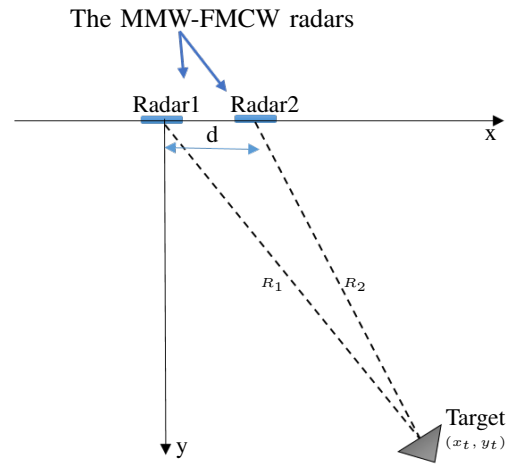
Following the CFAR implementation we will then obtain the radial distance of each target with respect to both radars. Based on Fig.~\ref{fig:Model}, for a point target located at $R_1$ and $R_2$ radial distance from radars number 1 and number 2, respectively, the x-y coordinate of the target based on the bilateration technique is calculated as
\begin{align}
\label{xy}
x_t = \frac{R^2_1 - R^2_2 + d^2}{2d}, \\ \nonumber
y_t = \sqrt{R^2_1 - x^2_1}.
\end{align}
In (\ref{xy}), the radar number 1 has been chosen as the origin of the coordinate system to simplify the equations.

In practice, however, we are dealing with multiple targets. In the case of multi-target scenario, we tune the element spacing $d$ with respect to the range resolution limit of the radar to avoid ghost targets.
We further implement a multi-target tracking block \cite{BarShalom_1,BarShalom_2} to help with removing the ghost targets and to provide smooth tracks for the targets.

\begin{figure}
\psfrag{x [mm]}[][]{$x$ [mm]}
\centerline{
\includegraphics[height=5cm,width=7cm]{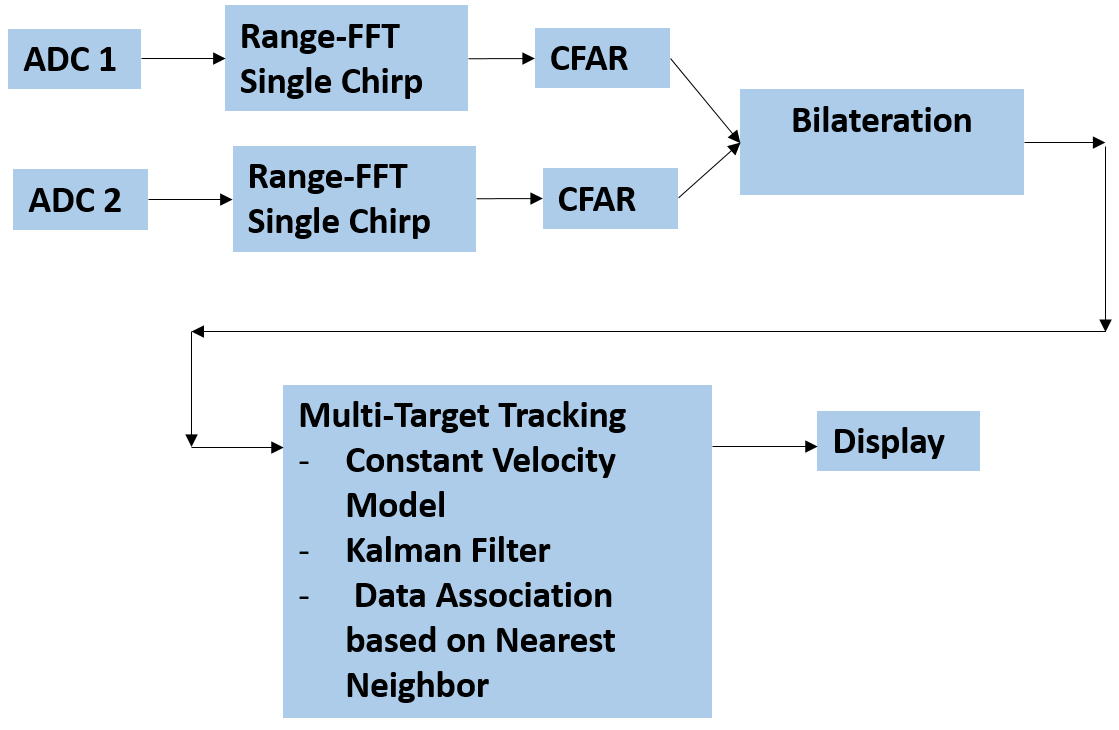}
}
\vspace*{0.2in}
\caption{The flowchart of the algorithm which shows the signal and data processing blocks.
\label{fig:DSP}}
\end{figure}

Since both the measurement and the motion model are linear, therefore, we use Kalman filter for tacking \cite{Mahafza, BarShalom_1,BarShalom_2}.
Moreover, we choose constant velocity model for the target motion. Also, we use Global Nearest Neighbor (GNN) method \cite{BarShalom_1,BarShalom_2} for the data association block.
In Fig.~\ref{fig:DSP}, we have presented the flowchart of the algorithm which includes both the signal processing and the data processing blocks that are necessary for the 2D localization of the targets.
\section{Experimental Results}\label{Experimental Results}
\begin{figure}[htb]
\centering
\begin{tikzpicture}
  \node (img1)  {\includegraphics[scale=0.7]{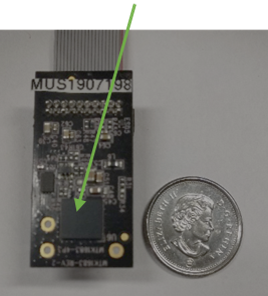}};
  \node[above=of img1, node distance=0cm, xshift=-0.4cm, yshift=-1.3cm,font=\color{black}] {The radar chip};
\end{tikzpicture}
\caption{The single MMW-FMCW radar from MediaTek company.}
\label{fig:radar}
\end{figure}

In this section, we present our experimental results. The single channel MMW-FMCW radar is from MediaTek company and has been shown in Fig.~\ref{fig:radar}.
We have chosen the same parameters for both radars. The center frequency of the radar is $\rm 79 \;GHz$ and the bandwidth has been set to $\rm 3.49 \; GHz$ which results in $\delta_R = 4.3\; cm$ resolution in the range direction. The chirp time is $\rm 68.8 \; \mu s$. We have set the maximum range of the radar to $\rm 5.5 \; m$. We receive $\rm 128$ samples per each chirp which means $\rm N=128$.

\begin{figure}[htb]
\centering
\begin{tikzpicture}
  \node (img1)  {\includegraphics[scale=0.7]{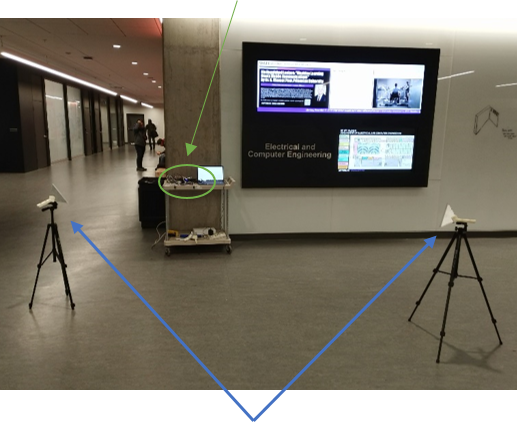}};
  \node[above=of img1, node distance=0cm, xshift=-0.4cm, yshift=-1.1cm,font=\color{black}] {The MMW-FMCW radars};
   \node[above=of img1, node distance=0cm, xshift=-0.4cm, yshift=-6.5cm,font=\color{black}] {The $\rm 20 \; dbsm$ corner reflectors};
\end{tikzpicture}
\caption{The test setup with two $\rm 20 \; dbsm$ corner reflectors and two MMW-FMCW radars.}
\label{fig:Setup_1}
\end{figure}

\begin{figure}[htb]
\begin{tikzpicture}
\node (img1){\includegraphics[scale=0.4]{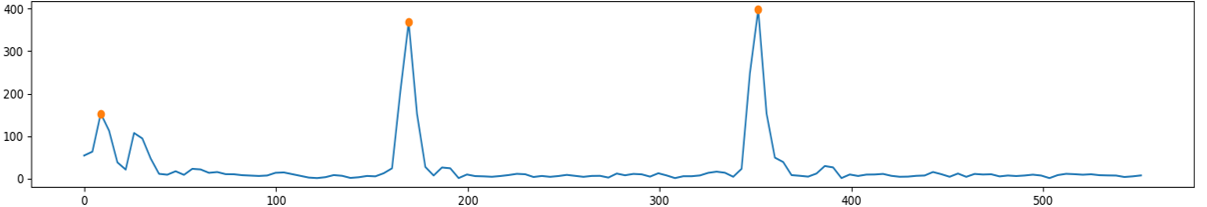}};
\node[below =of img1, node distance=0cm, xshift=0cm, yshift=1.2cm,font=\color{black}] {\tiny Radial Distance [cm]};
\node[below =of img1, node distance=0cm, xshift=0cm, yshift=1cm,font=\color{black}] {\small (a)};
\node[right=of img1, node distance=0cm,  xshift=-3.3cm, yshift=0.5cm,font=\color{black}] {\tiny Range profile for Radar1};
\node[below  = 0.4cm of img1](img2){\includegraphics[scale=0.4]{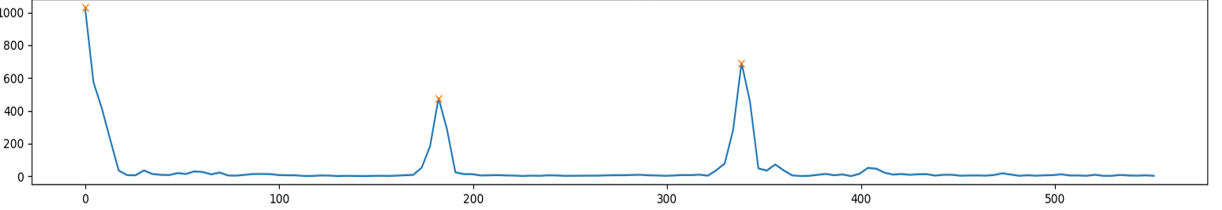}};
\node[below=of img2, node distance=0cm, xshift=0cm, yshift=1.2cm,font=\color{black}] {\tiny Radial Distance [cm]};
\node[below =of img2, node distance=0cm, xshift=0cm, yshift=1cm,font=\color{black}] {\small (b)};
\node[right=of img2, node distance=0cm,  xshift=-3.3cm, yshift=0.5cm,font=\color{black}] {\tiny Range profile for Radar2};
\node[below  = 0.4cm of img2](img3){\includegraphics[scale=0.4]{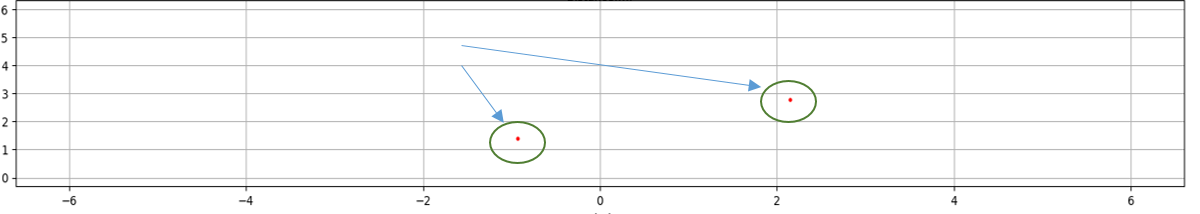}};
\node[below=of img3, node distance=0cm, xshift=0cm, yshift=1.2cm,font=\color{black}] {\tiny x[m]};
\node[below =of img3, node distance=0cm, xshift=0cm, yshift=1cm,font=\color{black}] {\small (c)};
\node[left=of img3, node distance=0cm, xshift=1.2cm, yshift=0.1cm,font=\color{black}] {\tiny y[m]};
\node[right=of img3, node distance=0cm,  xshift=-8.2cm, yshift=0.4cm,font=\color{black}] {\tiny The x-y position of targets};
\end{tikzpicture}
\caption{a) the range profile of radar1, b) the range profile of radar2, c) the result of bilateration based target localization for two $\rm 20 \; dbsm$ corner reflectors.}
\label{fig:Result_1}
\end{figure}
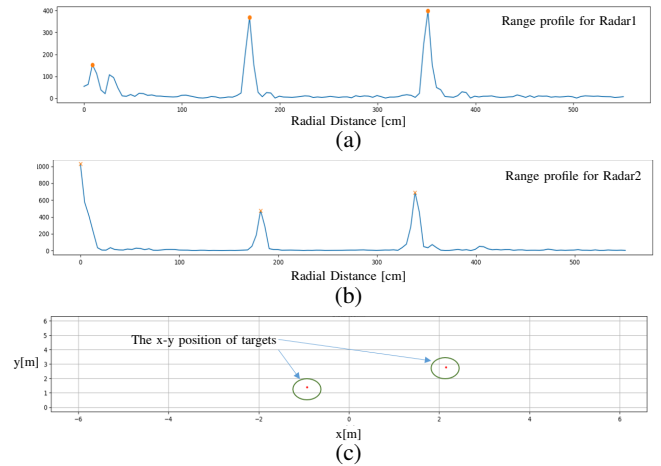

Fig.~\ref{fig:Setup_1} shows the test setup based on two MMW-FMCW radars. We have used two $\rm 20 \; dbsm$ corner reflectors as targets. The space between the two radars has been set to $\rm 20\;cm$. The value that is chosen for the element spacing plays an important role in dealing with ghost targets when bilateration method is used for target localization. We have also tuned the rate of birth and death of the targets in the multi-target tracking block to remove the ghost targets and to create a smooth track for each existing target.
Fig.~\ref{fig:Result_1} illustrates the result of the experiment performed based on the test setup shown in Fig.~\ref{fig:Setup_1}.

Fig.~\ref{fig:Result_1}-(a) and Fig.~\ref{fig:Result_1}-(b) depict the range profile and the output of the CFAR block, including the pruning process, for both the number 1 and the number 2 radars, respectively. The output of the CFAR has been shown as red dots in Fig.~\ref{fig:Result_1}-(a) and red crosses in Fig.~\ref{fig:Result_1}-(b). We have implemented the cell averaging CFAR for both radars.
The reflections very close to the radar are not from real targets and we have not sent them to the bilateration block.
After detecting the targets by both sensors, the output will then be sent to the bilateration block to calculate the x-y coordinates of the targets. The next step is to send the information about the  x-y coordinates of the targets to the multi-target tracking unit.
The final result has been shown in Fig.~\ref{fig:Result_1}-(c) which illustrates the x-y coordinates of the two corner reflectors.

For the multi-target tracking block we have used constant velocity model. Moreover, since both the measurement model and the dynamic model for the targets are linear with additive Gaussian noise, thus we have used Kalman filter. For the data association unit, global nearest neighbor method has been chosen. We have removed the ghost targets that appear in bilateration method by tuning the distance between the two radars relative to the range resolution as well as by tuning the rate of birth and death in the multi-target tracking block.

\begin{figure}[htb]
\centering
\begin{tikzpicture}
  \node (img1){\includegraphics[scale=0.7]{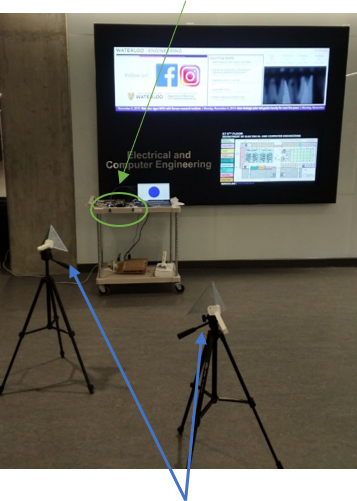}};
  \node[above=of img1, node distance=0cm, xshift=-0.2cm, yshift=-1.1cm,font=\color{black}] {The MMW-FMCW radars};
   \node[below=of img1, node distance=0cm, xshift=-0.4cm, yshift=1.3cm,font=\color{black}] {The $\rm 20 \; dbsm$ corner reflectors};
   \node[below =of img1, node distance=0cm, xshift=0cm, yshift=0.8cm,font=\color{black}] {\small (a)};

  \node[below  = 1cm of img1](img2){\includegraphics[scale=0.5]{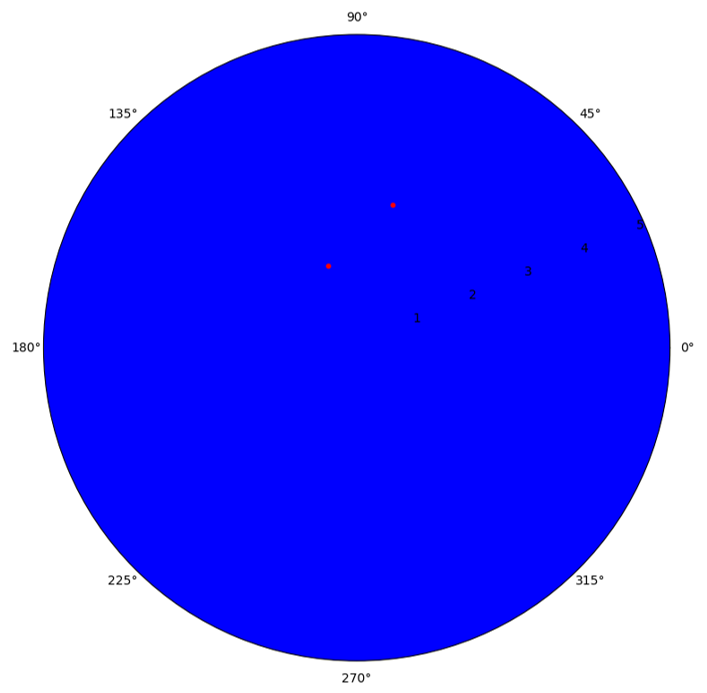}};
  \node[below =of img2, node distance=0cm, xshift=0cm, yshift=1cm,font=\color{black}] {\small (b)};

\end{tikzpicture}
\caption{a) the test setup for two $\rm 20 \; dbsm$ corner reflectors, b) the output of the system in polar coordinate system.}
\label{fig:Result_2}
\end{figure}
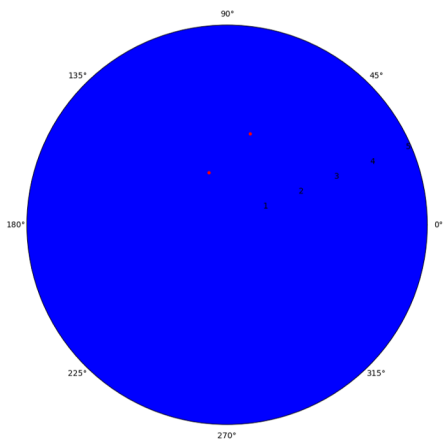

We have conducted another experiment using the same $\rm 20 \; dbsm$ corner reflectors and have illustrated the result in Fig.~\ref{fig:Result_2}.  The test setup has been depicted in Fig.~\ref{fig:Result_2}-(a) and the output of the system has been presented in polar coordinate system in Fig.~\ref{fig:Result_2}-(b).
\section{conclusion}
In this paper, we presented a 2D target localization method based on the bilateration technique using two low cost and compact MMW-FMCW radars operating at $\rm 79 \; GHz$.
\section{Acknowledgment}
The authors would like to express their appreciation to MediaTek company for funding the program and for providing them with the MMW-FMCW radars.

\bibliographystyle{IEEEtran}
\bibliography{Biblio}

\begin{thebibliography}{1}
\providecommand{\url}[1]{#1}
\csname url@samestyle\endcsname
\providecommand{\newblock}{\relax}
\providecommand{\bibinfo}[2]{#2}
\providecommand{\BIBentrySTDinterwordspacing}{\spaceskip=0pt\relax}
\providecommand{\BIBentryALTinterwordstretchfactor}{4}
\providecommand{\BIBentryALTinterwordspacing}{\spaceskip=\fontdimen2\font plus
\BIBentryALTinterwordstretchfactor\fontdimen3\font minus
  \fontdimen4\font\relax}
\providecommand{\BIBforeignlanguage}[2]{{%
\expandafter\ifx\csname l@#1\endcsname\relax
\typeout{** WARNING: IEEEtran.bst: No hyphenation pattern has been}%
\typeout{** loaded for the language `#1'. Using the pattern for}%
\typeout{** the default language instead.}%
\else
\language=\csname l@#1\endcsname
\fi
#2}}
\providecommand{\BIBdecl}{\relax}
\BIBdecl

\bibitem{Shahrokh}
S.~{Hamidi}, M.~{Nezhad-Ahmadi}, and S.~{Safavi-Naeini}, ``{TDM} based virtual
  {FMCW} {MIMO} radar imaging at 79{GHz},'' \emph{IEEE 18th International
  Symposium on Antenna Technology and Applied Electromagnetics {ANTEM}}, pp.
  1--2, Aug. 2018.

\bibitem{24GHzTDM}
M.~Harter, T.~Schipper, L.~Zwirello, A.~Ziroff, and T.~Zwick, ``{24GHz} digital
  beamforming radar with t-shaped antenna array for three-dimensional object
  detection,'' \emph{International Journal of Microwave and Wireless
  Technologies}, vol.~4, no.~3, 2012.

\bibitem{77GHzTDM}
R.~Feger, C.~Pfeffer, C.~M. Schmid, M.~J. Lang, Z.~Tong, and A.~Stelzer, ``A
  77-{GHz} {FMCW} {MIMO} radar based on loosely coupled stations,'' \emph{7th
  German Microwave Conference}, pp. 1--4, Mar. 2012.

\bibitem{TDMcalibration}
J.~Guetlein, A.~Kirschner, and J.~Detlefsen, ``Calibration strategy for a {TDM}
  {FMCW} {MIMO} radar system,'' \emph{IEEE International Conference on
  Microwaves, Communications, Antennas and Electronic Systems (COMCAS 2013)},
  pp. 1--5, Oct. 2013.

\bibitem{FDMMIMO}
M.~K. A.~Zwanetski and H.~Rohling, ``Waveform design for {FMCW MIMO} radar
  based on frequency division,'' \emph{IRS 2013, 14th International Radar
  Symposium, Dresden, DE}, pp. 19--21, Jun. 2013.

\bibitem{TDMmotioncompensation}
C.~M. Schmid, R.~Feger, C.~Pfeffer, and A.~Stelzer, ``Motion compensation and
  efficient array design for {TDMA} {FMCW} {MIMO} radar systems,'' \emph{6th
  European Conference on Antennas and Propagation (EUCAP)}, pp. 1746--1750,
  Mar. 2012.

\bibitem{Mahafza}
B.~R. Mahafza, \emph{Radar Signal Analysis and Processing Using
  {MATLAB}}.\hskip 1em plus 0.5em minus 0.4em\relax Chapman and Hall/CRC, 2008.

\bibitem{BarShalom_1}
Y.~Bar-Shalom, T.~Fortmann, P.~Howlett, and A.~Torokhti, \emph{Tracking And
  Data Association by Yaakov BarShalom and Thomas E Fortmann}.\hskip 1em plus
  0.5em minus 0.4em\relax Elsevier Science, 1988.

\bibitem{BarShalom_2}
Y.~Bar-Shalom and W.~Blair, \emph{Multitarget-multisensor Tracking:
  Applications and Advances}.\hskip 1em plus 0.5em minus 0.4em\relax Artech
  House, 1990.

\end{thebibliography}

\end{document}